\documentclass[twocolumn,aps,superscriptaddress]{revtex4}
\usepackage{amssymb}
\usepackage{amsmath}
\usepackage{graphicx}
\usepackage{epstopdf}
\usepackage[normalem]{ulem}
\usepackage[dvips]{color}
\usepackage{multirow}

\renewcommand{\sout}{\bgroup \color{red} \ULdepth=-.5ex \ULset}

\begin{document}

\title{Nuclear matter properties at finite temperatures from effective interactions}
\author{Jun Xu\footnote{xujun@zjlab.org.cn}}
\affiliation{Shanghai Advanced Research Institute, Chinese Academy of Sciences, Shanghai 201210, China}
\affiliation{Shanghai Institute of Applied Physics, Chinese Academy of Sciences, Shanghai 201800, China}
\author{Arianna Carbone\footnote{acarbone@ectstar.eu}}
\affiliation{European Centre for Theoretical Studies in Nuclear Physics and Related Areas (ECT*) and Fondazione Bruno Kessler, Strada delle Tabarelle 286, I-38123 Villazzano (TN), Italy}
%\author{Jeremy W. Holt\footnote{holt@physics.tamu.edu}}
%\affiliation{Cyclotron Institute and Department of Physics and Astronomy, Texas A$\&$M University, College Station, TX 77843, USA}
\author{Zhen Zhang\footnote{zhangzh275@mail.sysu.edu.cn}}
\affiliation{Sino-French Institute of Nuclear Engineering and Technology, Sun Yat-Sen University, Zhuhai 519082, China}
\author{Che Ming Ko\footnote{ko@comp.tamu.edu}}
\affiliation{Cyclotron Institute and Department of Physics and Astronomy, Texas A$\&$M University, College Station, TX 77843, USA}

\date{\today}

\begin{abstract}
We study if commonly used nucleon-nucleon effective interactions, obtained from fitting the properties of cold nuclear matter and of finite nuclei, can properly describe the hot dense nuclear matter produced in intermediate-energy heavy-ion collisions. We use two representative effective interactions, i.e., an improved isospin- and momentum-dependent interaction with its isovector part calibrated by the results from the \emph{ab initio} non-perturbative self-consistent Green's function (SCGF) approach with chiral forces, and a Skyme-type interaction fitted to the equation of state of cold nuclear matter from chiral effective many-body perturbation theory and the binding energy of finite nuclei. In the mean-field approximation, we evaluate the equation of state and the single-nucleon potential for nuclear matter at finite temperatures and compare them to those from the SCGF approach. We find that the improved isospin- and momentum-dependent interaction reproduces reasonably well the SCGF results due to its weaker momentum dependence of the mean-field potential than in the Skyrme-type interaction. Our study thus indicates that effective interactions with the correct momentum dependence of the mean-filed potential can properly describe the properties of hot dense nuclear matter and are thus suitable for use in transport models to study heavy-ion collisions at intermediate energies.
\end{abstract}

\maketitle

\section{Introduction}
\label{introduction}

One of the main motivations for pursuing experiments on heavy-ion collisions at intermediate energies is to study the equation of state (EOS) of nuclear matter.
%, i.e., the density dependence of its binding energy.
Its knowledge is essential for understanding the properties of systems ranging from finite nuclei~\cite{Ste05,Che10} to neutron stars~\cite{Lat16,Heb13} as well as the gravitational-wave signal from neutron star mergers~\cite{Baus12,Rezz16,Lim18}. Because of the complexities of heavy-ion collision dynamics, transport models have been indispensable tools to extract the information on the nuclear EOS, particularly at high densities that exist during the early stage of the collisions, from various observables measured in
experiments~\cite{Dan02,Aic85,Fuc01,Li08}.  In transport models, which are based on either the Boltzmann-Uehling-Uhlenbeck equation~\cite{Ber88} or the quantum molecular dynamics~\cite{Aic91}, the time evolution of nucleon phase-space distribution functions in a heavy-ion collision is determined by both the mean-field potential acting on nucleons and their scatterings. The nucleon mean-field potential is usually obtained from nucleon-nucleon (NN) effective interactions that are constructed from fitting the properties of cold nuclear matter and of finite nuclei. Thus, the mean-field potential does not include explicitly the temperature effect on the NN effective interactions in the nuclear medium, which is needed to describe the hot nuclear matter produced in heavy-ion collisions. On the other hand, the mean-field potential extracted from comparing results of transport models with the experimental data is at finite temperatures, so it cannot be assumed {\it a priori} to be related to the nuclear EOS at zero temperature based on the quasi-particle model and be used to safely constrain the cold nuclear matter properties.

The above assumption can be justified if the mean-field potential obtained from NN effective interactions and used in transport models can also fit the finite-temperature single-nucleon potential obtained from microscopic calculations, such as that based on the self-consistent Green's function (SCGF) approach~\cite{Carbone2014} or the many-body perturbation theory~\cite{Wel15,Wel16} employing chiral nuclear forces~\cite{Epelbaum2009,Machleidt2011}. In the present study, we choose NN effective interactions that correspond to two energy-density functionals based on the Hartree-Fock calculations. One is obtained from an improved isospin- and momentum-dependent interaction (ImMDI) model~\cite{Das03,Xu15}, which is constructed from fitting cold nuclear matter properties at saturation density and the empirical nucleon optical potential. The other is the Skyrme-Hartree-Fock (SHF) model~\cite{Cha97,Che10,Dut12} using the Sk$\chi$m$^*$ force, which is constructed by fitting the properties of cold nuclear matter from chiral effective many-body perturbation theory ($\chi$EMBPT) and the binding energies of finite nuclei~\cite{Zha18}. The properties of cold neutron matter from the SCGF approach were further used to constrain the isovector part of the ImMDI model, and the new parametrization of this effective interaction is dubbed as ImMDI-GF. The ImMDI model and the Sk$\chi$m$^*$ interaction, which have been used in transport models to study heavy-ion collisions at intermediate energies~\cite{Xu18,Zha18a}, are then evaluated in the non-relativistic mean-field approximation to obtain the properties of symmetric nuclear matter (SNM) and pure neutron matter (PNM) at finite temperatures. These results are compared with those from the SCGF approach~\cite{Carbone2014,Carbone2018} and the $\chi$EMBPT approach~\cite{Wel15,Wel16} within their theoretical uncertainties, which are mainly due to the variation of the high-momentum cutoff in nuclear interactions and the three-body forces included in these microscopic studies.

The above comparison shows that the effective interaction of proper momentum dependence in its mean-field potential is able to reproduce reasonably well the properties of nuclear matter at finite temperatures from microscopic calculations using chiral forces. In the non-relativistic framework as discussed in the present study, the momentum dependence of the mean-field potential, or the related nucleon effective k-mass, is from the Fock contribution of the non-local effective interaction. This is different from the relativistic approach, where the momentum dependence of the Sch\"ordinger-equivalent potential originates from the nucleon Dirac mass through its coupling to a scalar meson. The nonlocality of the relativistic interaction in time may also lead to the energy dependence of the mean-field potential, and can be characterized by the so-called nucleon effective E-mass. For detailed discussions on the nucleon effective mass as well as the momentum dependence of the nuclear mean-field potential, we refer the reader to Ref.~\cite{Jam89}.

The remaining part of the paper is organized as follows. Section~\ref{framework} gives the details on the theoretical framework for the ImMDI model and the SHF model as well as the SCGF approach. In Sec.~\ref{results}, we compare and discuss the results for the occupation probabilities, the entropies and the heat capacities, the EOSs, and the mean-field potentials for SNM and PNM obtained from these different approaches. Finally, a summary is given in Sec.~\ref{summary}.

\section{Theoretical framework}
\label{framework}

\subsection{Effective interactions in Hartree-Fock calculations}
\label{EIHF}

The effective interaction between two nucleons at coordinates $\vec{r}_1$ and $\vec{r}_2$ in the ImMDI model includes a zero-range density-dependent term and a Yukawa-type finite-range term~\cite{Xu10}, i.e.,
\begin{eqnarray}
&&v_{\rm ImMDI}(\vec{r}_1,\vec{r}_2)\notag\\
&&= \frac{1}{6}t_3(1+x_3 P_\sigma)\rho^\alpha\left(\frac{\vec{r}_1+\vec{r}_2}{2}\right)\delta(\vec{r}_1-\vec{r}_2) \notag\\
&&+ (W+B P_\sigma - H P_\tau - M P_\sigma P_\tau) \frac{e^{-\mu |\vec{r}_1-\vec{r}_2}|}{|\vec{r}_1-\vec{r}_2|},\notag\\
\label{MDIv}
\end{eqnarray}
where $\rho$ is the nucleon number density, $P_\sigma$ and $P_\tau$ are the spin and isospin exchange operators, respectively, and $t_3$, $x_3$, $\alpha$, $W$, $B$, $H$, $M$, and $\mu$ are parameters.

The standard Skyrme interaction~\cite{Cha97} without the spin-orbit coupling has the form of
\begin{eqnarray}
&&v_{\rm SHF}(\vec{r}_1,\vec{r}_2) \notag\\
&&= t_0(1+x_0 P_\sigma)\delta(\vec{r}_1-\vec{r}_2)\notag\\
&&+ \frac{1}{2}t_1(1+x_1P_\sigma)[\vec{k'}^2 \delta(\vec{r}_1-\vec{r}_2) + \delta(\vec{r}_1-\vec{r}_2) \vec{k}^2]\notag\\
&&+ t_2(1+x_2P_\sigma)\vec{k'} \cdot \delta(\vec{r}_1-\vec{r}_2) \vec{k}\notag\\
&&+\frac{1}{6}t_3(1+x_3 P_\sigma)\rho^\alpha\left(\frac{\vec{r}_1+\vec{r}_2}{2}\right)\delta(\vec{r}_1-\vec{r}_2), \notag\\
\label{SHFv}
\end{eqnarray}
where $\vec{k}=\frac{1}{2i}(\nabla_1-\nabla_2)$ is the relative momentum operator acting on the right-hand side, $\vec{k'}$ is the complex conjugate of $\vec{k}$ acting on the left-hand side, and $t_0$, $x_0$, $t_1$, $x_1$, $t_2$, $x_2$, $t_3$, $x_3$, and $\alpha$ are parameters.

In the Hartree-Fock approach, the total potential energy of nuclear matter is calculated according to
\begin{equation}\label{vV}
E_p = \frac{1}{2} \sum_{i,j} <ij|v (1-P_r P_\sigma P_\tau)|ij>,
\end{equation}
where $P_r$ is the space exchange operator, $|i(j)>$ is the quantum state of $i(j)$th nucleon, and $v$ is the NN effective interaction.

The potential energy density from the ImMDI model is then given by~\cite{Xu15}
\begin{eqnarray}
V_{\rm ImMDI} &=&\frac{A_{u}\rho _{n}\rho _{p}}{\rho _{0}}+\frac{A_{l}}{%
 	2\rho _{0}}(\rho _{n}^{2}+\rho _{p}^{2})+\frac{B}{\sigma+1}\frac{\rho^{\sigma +1}}{\rho _{0}^{\sigma }}  \notag \\
 & &\times (1-x\delta ^{2})+\frac{1}{\rho _{0}}\sum_{q ,q^{\prime}}C_{q ,q ^{\prime }}  \notag \\
 & &\times \int \int d^{3}pd^{3}p^{\prime }\frac{f_{q }(\vec{r}, \vec{p}%
 	)f_{q ^{\prime }}(\vec{r}, \vec{p}^{\prime })}{1+(\vec{p}-\vec{p}^{\prime})^{2}/\Lambda ^{2}}, \label{MDIEp}
\end{eqnarray}
where $\rho_n$ and $\rho_p$ are the neutron and proton number densities, respectively, $\rho_{0}=0.16$ fm$^{-3}$ is a constant density, $\delta =(\rho _{n}-\rho _{p})/\rho$ is the isospin asymmetry of nuclear matter with $\rho=\rho_n+\rho_p$, and $f_{q}(\vec{r}, \vec{p})$ is the nucleon phase-space distribution function obtained from the Wigner transformation of its density matrix\ with $q=1$ for neutrons and $-1$ for protons. For the detailed derivation of the above expression, we refer the reader to Ref.~\cite{Xu10}, where the relation between values of the parameter sets $(t_3,x_3,\alpha,W,B,H,M,\alpha)$ and $(A_u,A_l,B,C_{q,-q},C_{q,q},\Lambda,\sigma,x)$ can be found.
In Ref.~\cite{Xu15}, an optimized parameter set
$(A_0,B,C_{l0},C_{u0},\Lambda,\sigma,x,y)$ was introduced by using following relations:
\begin{eqnarray}
 A_{l}(x,y)&=&A_{0} + y + x\frac{2B}{\sigma +1},   \label{AlImMDI}\\
 A_{u}(x,y)&=&A_{0} - y - x\frac{2B}{\sigma +1},   \label{AuImMDI}\\
 C_{q,q}(y)&=&C_{l0} - 2(y-2z)\frac{p^2_{f0}}{\Lambda^2\ln [(4 p^2_{f0} + \Lambda^2)/\Lambda^2]},   \label{ClImMDI}\\
 C_{q,-q}(y)&=&C_{u0} + 2(y-2z)\frac{p^2_{f0}}{\Lambda^2\ln[(4 p^2_{f0} + \Lambda^2)/\Lambda^2]},   \label{CuImMDI}
\end{eqnarray}
where $p_{f0}=\hbar(3\pi^{2}\rho_0/2)^{1/3}$ is the nucleon Fermi momentum in SNM at $\rho_0$. The number of independent parameters in the new set is the same as before. The parameters $x$, $y$, and $z$ then characterize the slope parameter of the symmetry energy, the momentum dependence of the symmetry potential, and the symmetry energy at $\rho_0$, respectively.

The potential energy density in uniform nuclear matter for the SHF model is given by
\begin{eqnarray}
V_{\rm SHF} &=& t_{0}[(2+x_{0})\rho ^{2}-(2x_{0}+1)(\rho _{p}^{2}+\rho
_{n}^{2})]/4 \notag\\
&+&[t_{1}(2+x_{1})+t_{2}(2+x_{2})]\tau \rho /8 \notag\\
&+&[t_{2}(2x_{2}+1)-t_{1}(2x_{1}+1)](\tau _{n}\rho
_{n}+\tau _{p}\rho _{p})/8\notag\\
&+&t_{3}\rho ^{\sigma }[(2+x_{3})\rho ^{2}-(2x_{3}+1)(\rho
_{p}^{2}+\rho _{n}^{2})]/24,
\end{eqnarray}
where $\tau=\sum_{q}\tau_q$ is the total kinetic density with $\tau_{q}=\int p^2 f_{q}(\vec{r}, \vec{p}) d^3p /(2\pi)^3$ being that for nucleons with isospin $q$.

Through the variational principle, the mean-field potential for a nucleon with momentum $\vec{p}$ and isospin $q$ in the asymmetric nuclear matter of isospin asymmetry $\delta$ and nucleon number density $\rho$ from the ImMDI model can be expressed as~\cite{Xu15}
\begin{eqnarray}
 U_{\rm ImMDI} &=&A_{u}\frac{\rho _{-q }}{\rho _{0}}%
 +A_{l}\frac{\rho _{q }}{\rho _{0}}  \notag \\
 & & +B\left(\frac{\rho }{\rho _{0}}\right)^{\sigma }(1-x\delta ^{2}) - 4qx\frac{B}{%
 	\sigma +1}\frac{\rho ^{\sigma -1}}{\rho _{0}^{\sigma }}\delta \rho
 _{-q }
 \notag \\
 & & +\frac{2C_{q,q}}{\rho _{0}}\int d^{3}p^{\prime }\frac{f_{q }(%
 	\vec{r}, \vec{p}^{\prime })}{1+(\vec{p}-\vec{p}^{\prime })^{2}/\Lambda ^{2}}
 \notag \\
 & & +\frac{2C_{q,-q}}{\rho _{0}}\int d^{3}p^{\prime }\frac{f_{-q }(%
 	\vec{r}, \vec{p}^{\prime })}{1+(\vec{p}-\vec{p}^{\prime })^{2}/\Lambda ^{2}}.
 \label{MDIU}
\end{eqnarray}%

Similarly, the mean-field potential in the standard SHF model can be expressed as
\begin{eqnarray}\label{SHFU}
U_{\rm SHF} &=& \frac{p^2}{2m_q^*} - \frac{p^2}{2m} + t_0\left(1+\frac{x_0}{2}\right)\rho - t_0\left(\frac{1}{2}+x_0\right)\rho_q \notag\\
&+&\frac{1}{4}\left[t_1\left(1+\frac{x_1}{2}\right)+t_2\left(1+\frac{x_2}{2}\right)\right]\tau \notag\\
&-&\frac{1}{4}\left[t_1\left(\frac{1}{2}+x_1\right)-t_2\left(\frac{1}{2}+x_2\right)\right]\tau_q \notag\\
&+&\frac{1}{12}t_3\rho^\alpha\left[(2+\alpha)\left(1+\frac{x_3}{2}\right)\rho-\left(1+2x_3\right)\rho_q\right.\notag\\
&-&\left.\alpha\left(\frac{1}{2}+x_3\right)\frac{\rho_n^2+\rho_p^2}{\rho}\right],
\end{eqnarray}
where the effective mass $m_q^*$ is given by
\begin{eqnarray}
\frac{1}{2m_q^*} &=& \frac{1}{2m} + \frac{1}{4}\left[t_1\left(1+\frac{x_1}{2}\right)+t_2\left(1+\frac{x_2}{2}\right)\right]\rho \notag\\
&-& \frac{1}{4} \left[t_1\left(\frac{1}{2}+x_1\right)-t_2\left(\frac{1}{2}+x_2\right)\right]\rho_q.
\end{eqnarray}
In the above, $m$ is the bare nucleon mass, which is taken to be the same for neutron and proton.

It is well known that the Fock exchange contribution from the finite-range term in Eq.~\eqref{MDIv} in the ImMDI model leads to a momentum dependence in the mean-field potential [Eq.~\eqref{MDIU}], with its form given by the Fourier transform of the Yukawa-type finite-range interaction. Historically, such a momentum dependence was developed from a momentum-dependent Yukawa-type interaction~\cite{Gal87,Ber88}, in order to reproduce the measured nucleon transverse flow in heavy-ion collisions with a reasonable nuclear matter incompressibility. For the Skyrme interaction, it only contains the lowest-order non-local term and can be considered as a low-momentum expansion of the finite-range interaction. Consequently, the mean-field potential from the SHF model with a quadratic momentum dependence [Eq.~\eqref{SHFU}] is valid only at low momenta, and it becomes less valid with increasing nucleon momentum and also in nuclear medium of higher density or larger Fermi momentum.

In a uniform and thermalized nuclear medium, the nucleon phase-space distribution function $f_q(\vec{r},\vec{p})$ has the Fermi-Dirac form, i.e., $f_q(\vec{r},\vec{p})=2/\{\exp[(p^2/2m+U_q-\mu_q)/T]+1\}$, where $T$, $\mu_q$ and $U_q$ are, respectively, the temperature, the chemical potential, and the mean-field potential. Since the mean-field potential $U_q$ from the ImMDI model depends on the phase-space distribution function $f_q(\vec{r},\vec{p})$, calculations of nuclear matter properties at finite temperatures~\cite{Xu07} need to be carried out self-consistently using the iteration method. The calculation in the SHF model for nuclear matter at finite temperatures is simpler, since the effective mass of a nucleon depends only on density and not on its momentum.

The total energy per nucleon $E/A$, which consists of both the potential and kinetic energy contributions, is given by
\begin{equation}
E/A = E_p + E_k = \frac{V}{\rho} + \sum_q \frac{\tau_q}{2m\rho}.
\end{equation}

\subsection{Green's function approach using chiral forces}

The SCGF method is a nonperturbative many-body approach based on the calculation of the dressed nucleon propagator, i.e., its Green's function $G$~\cite{Dickhoff2004}. The single-particle propagator provides access to microscopic properties of the many-body system, such as the nucleon spectral function or momentum distribution, and also to bulk thermodynamical quantities, such as internal energy, entropy, pressure, etc. Within this approach, the dressed propagator $G$ is obtained via the iterative solution of the Dyson's equation
\begin{equation}
G({\bf p},\omega)=G_{0}({\bf p},\omega)+G_{0}({\bf p},\omega)\Sigma^\star({\bf p},\omega)G({\bf p},\omega)\,,
\label{eq:dyson}
\end{equation}
where a nonperturbative self-energy $\Sigma^\star({\bf p},\omega)$ is employed, with ${\bf p}$ and $\omega$ being the single-particle momentum and energy. The self-energy is obtained within the so-called \emph{ladder approximation}, where an infinite resummation of particle-particle and hole-hole intermediate states is considered. Hence the method is nonperturbative and self-consistent, providing a fully correlated description of the many-body system beyond the mean-field level~\cite{Rios2009}. In recent years the SCGF approach has been extended to consistently include two- and three-body forces~\cite{Carbone2013}. Within this improved approach, the energy per nucleon can be obtained via an extended energy sum rule that reads~\cite{Carbone2013}
\begin{equation}
\frac{E}{A}=\frac{\nu}{\rho}\int\frac{{\rm d}{\bf p}}{(2\pi)^3}\int\frac{{\rm d}\omega}{2\pi}\frac{1}{2}\Big(\frac{p^2}{2m}+\omega\Big)\mathcal{A}({\bf p},\omega)f(\omega)-\frac{1}{2}\langle \hat W \rangle\,.
\label{eq:energy}
\end{equation}
In the above, $\nu=2$ for PNM and 4 for SNM is the nucleon spin-isospin degeneracy, $\rho$ is again the total nucleon number density, $\mathcal{A}({\bf p},\omega)$ is the spectral function, $f(\omega)$ is the Fermi-Dirac distribution; and $\langle \hat W \rangle$ is the expectation value of the three-body operator. The spectral function $\mathcal{A}({\bf p},\omega)$, which enters the calculation of the energy sum rule, is directly connected with the single-particle propagator $G$, being proportional to its imaginary part~\cite{Dickhoff2004}. From the spectral function one has direct access to the nucleon momentum distribution
\begin{equation}
n({\bf p})=\int\frac{{\rm d}\omega}{2\pi}\mathcal{A}({\bf p},\omega)f(\omega).
\label{momdis}
\end{equation}
One can then evaluate the kinetic energy contribution $E_k$ to the energy per nucleon according to
\begin{equation}\label{kinetic}
E_k = \frac{\nu}{\rho}\int\frac{{\rm d}{\bf p}}{(2\pi)^3}\frac{p^2}{2m}n({\bf p}),
\end{equation}
as well as the potential energy contribution $E_p$ via subtraction of Eq.~\eqref{kinetic} from Eq.~\eqref{eq:energy}. For the nucleon energy spectrum, it is obtained by solving consistently the equation
\begin{equation}\label{epsilon}
\varepsilon({\bf p})=\frac{p^2}{2m}+{\rm Re}\Sigma^\star[{\bf p},\varepsilon({\bf p})].
\end{equation}
The second term on the right-hand side only selects the on-shell part of the real self-energy, and it is what corresponds to a mean-field potential. For further details on the calculation of the finite-temperature properties of infinite matter within the SCGF method, we refer the reader to Ref.~\cite{Rios2009}.

The extension of the SCGF method to include three-body forces paved the way to the possibility of using consistently nuclear interactions derived from the chiral effective field theory. These interactions, being derived from a low-energy effective theory of QCD, have a cutoff in momentum usually around $\sim500$ MeV/c. The high-energy physics, which is integrated out, is then encoded in low-energy constants, which need to be fitted to nuclear matter properties~\cite{Epelbaum2009,Machleidt2011}. Studies of the properties of infinite matter at both zero and finite temperatures have been presented within the SCGF method for several different chiral interactions~\cite{Carbone2014,Carbone2018}. In this work we make use of three different chiral interactions. These have been chosen because they predict reasonably well the empirical saturation properties of symmetric nuclear matter~\cite{Carboneunpub}.  We are then able to provide an error band on our theoretical results based on the nuclear interaction. Two of these interactions, i.e., 2.0/2.0(EM) and 2.0/2.5(EM), have been obtained by fitting the two-body part to nucleon-nucleon phase shifts and deuteron properties, while the three-body part has been constructed to reproduce the binding energy of tritons and the radius of alpha particles. The two-body part has been further softened with the similarity renormalization group technique to improve the convergency of many-body calculations, as detailed in Ref.~\cite{Hebeler2011}. The third interaction is called NNLOsat with the whole two- and three-body parts fitted consistently, and it can reproduce reasonably well the properties of light nuclei as well as those of medium-mass nuclei, such as the radii of carbon and oxygen isotopes~\cite{Ekstrom2015}.

%\subsection{Chiral effective field theory}

\section{Results and discussions}
\label{results}

%\subsection{Parametrization of the effective interactions}

In the following, we compare some properties of infinite nuclear matter obtained from the two effective interactions ImMDI-GF and Sk$\chi$m$^*$ using the Hartree-Fock approach to those from the chiral forces based on microscopic SCGF calculations, whose uncertainties mainly come from those in the three-body forces and the high-momentum cutoffs. As stated in the introduction, the ImMDI model is fitted to the empirical properties of cold SNM, which are approximately reproduced by the SCGF approach using the chiral forces. As an improvement of the ImMDI model, we adjust the parameters of its isovector part, i.e., $x$, $y$, and $z$, to reproduce the results from the SCGF approach for the properties of PNM at zero temperature, and this new parameter set is dubbed as ImMDI-GF. For the SHF energy density functional, the Sk$\chi$$m^*$ interaction used in the present study is constructed from fitting the EOS and nucleon effective masses of cold nuclear matter from the $\chi$EMBPT and the binding energies of finite nuclei~\cite{Zha18}. Details on the values of the parameters in ImMDI-GF and Sk$\chi$m$^*$ interactions as well as some of their predicted physical quantities are listed in Table.~\ref{para}.

\begin{table}\small
  \caption{Values of parameters and some physical quantities for ImMDI-GF and Sk$\chi$m$^*$, with $\rho_{sat}$ the saturation density, $E_0(\rho_{sat})$ the energy per nucleon at saturation density, $K_0$ the incompressibility, $U_0^\infty$ the mean-field potential for SNM at saturation density and infinitely large nucleon momentum, $m_s^*$ and $m_v^*$ the isoscalar and the isovector effective mass, $E_{sym}(\rho_{sat})$ and $L$ the value and the slope parameter of the symmetry energy at saturation density, and $G_S$ and $G_V$ the isoscalar and the isovector density gradient coefficient.}
    \begin{tabular}{|c | c|| c|c|}
    \hline
      & ImMDI-GF &  & Sk$\chi$m$^*$ \\
   \hline
    $A_0$ (MeV)     & -66.963 & $t_0$ (MeVfm$^{3}$) & -2260.7  \\
   \hline
    $B$ (MeV)     & 141.963 & $x_0$ & 0.327488  \\
   \hline
    $C_{u0}$ (MeV)    & -99.70 & $t_1$ (MeVfm$^{5}$) & 433.189 \\
   \hline
    $C_{l0}$ (MeV) & -60.49 & $x_1$ & -1.088968  \\
   \hline
    $\sigma$         & 1.2652 & $t_2$ (MeVfm$^{5}$) & 274.553 \\
   \hline
    $\Lambda$ ($p_{f0}$)    & 2.424 & $x_2$ & -1.822404  \\
   \hline
    $x$           & 0.5 & $t_3$ (MeVfm$^{3+3\alpha}$)& 12984.4 \\
   \hline
    $y$ (MeV)     & -60 & $x_3$ & 0.442900  \\
    \hline
    $z$ (MeV)     & -2.5 & $\alpha$ & 0.198029 \\
   \hline
   \hline
     $\rho_{sat}$ (fm$^{-3}$) & 0.16 & $\rho_{sat}$ (fm$^{-3}$) & 0.1651 \\
   \hline
     $E_0(\rho_{sat})$ (MeV) & -16 & $E_0(\rho_{sat})$ (MeV) & -16.07 \\
   \hline
     $K_0$ (MeV)  & 230 & $K_0$ (MeV) & 230.4 \\
   \hline
     $U_0^\infty$ (MeV)   & 75 & $U_0^\infty$ (MeV) & N/A \\
   \hline
     $m_s^*$ ($m$)  & 0.70 & $m_s^*$ ($m$) & 0.750 \\
   \hline
     $E_{sym}(\rho_{sat})$ (MeV)   & 30 & $E_{sym}(\rho_{sat})$ (MeV) & 30.94 \\
   \hline
     $L$ (MeV)        & 40 & $L$ (MeV) & 45.6 \\
   \hline
     $m_v^*$ ($m$)   & 0.59 & $m_v^*$ ($m$) & 0.694 \\
   \hline
        $G_S$ (MeVfm$^5$)  & N/A & $G_S$ (MeVfm$^5$) & 141.5 \\
   \hline
        $G_V$ (MeVfm$^5$)   & N/A & $G_V$ (MeVfm$^5$) & -70.5 \\
   \hline
    \end{tabular}
  \label{para}
\end{table}

%\subsection{Comparison of the EOS}

\subsection{Nucleon occupation probability}

\begin{figure}[htbp]
	\includegraphics[scale=0.35]{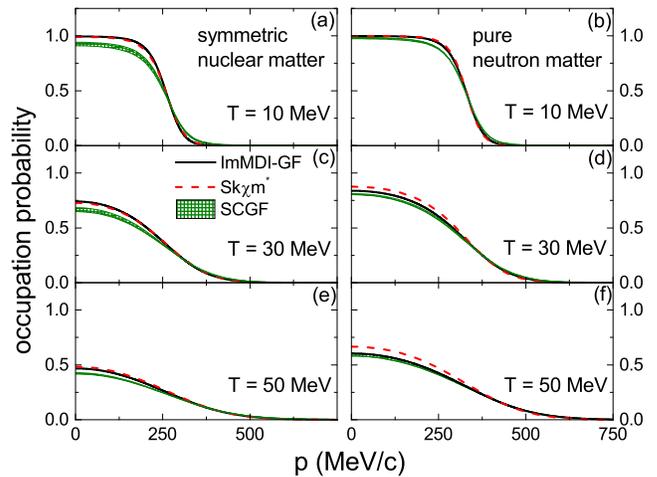}
	\caption{(Color	online) Nucleon occupation probability at $\rho_0=0.16$ fm$^{-3}$ as a function of nucleon momentum in symmetric nuclear matter (left) and pure neutron matter (right) at various temperatures from the ImMDI-GF, Sk$\chi$m$^*$, and SCGF calculations.}\label{f}
\end{figure}

We first show in Fig.~\ref{f} the nucleon occupation probability $n(p)$ at $\rho_0=0.16$ fm$^{-3}$ in both SNM and PNM at temperatures of 10, 30, and 50 MeV. For results from ImMDI-GF and Sk$\chi$m$^*$ denoted, respectively, by solid and dashed lines, the occupation probability is calculated according to the Fermi-Dirac distribution, i.e., $n(p)=1/\{\exp[(p^2/2m+U_q-\mu_q)/T]+1\}$. For results from the SCGF calculations, they are obtained from Eq.~\eqref{momdis}, which depends on the off-shell nucleon spectral function, and they are represented by shaded bands due to the uncertainties in this approach of using the three chiral forces 2.0/2.0(EM), 2.0/2.5(EM), and NNLOsat. It is worthy to point out that while the occupation probabilities at zero temperature are simply $\Theta(p_f-p)$ in the on-shell mean-field models as in the cases of ImMDI-GF and Sk$\chi$m$^*$, the sharp discontinuity at the Fermi momentum is smoothed in the off-shell SCGF calculations by correlations in the nuclear many-body system~\cite{Ding2016}.

For SNM, the occupation probabilities obtained from ImMDI-GF and Sk$\chi$m$^*$ differ from those from the SCGF approach at all temperatures, with the latter more depleted at low momenta due to the inclusion of correlation effects. For PNM, the occupation probabilities from ImMDI-GF and Sk$\chi$m$^*$ are similar at low temperatures but start to deviate with increasing temperature. In this case, the ImMDI-GF results are closer to the SCGF ones while those from Sk$\chi$m$^*$ remain higher at low momenta as temperature increases. This is likely due to the different momentum dependence of the mean-field potential in these approaches (see later Fig.~\ref{U_PNM}). A closer comparison to results from the SCGF approach shows that the nucleon occupation probability from Sk$\chi$m$^*$ is always larger at lower momenta and smaller at higher momenta. This is caused by the quadratic momentum dependence in its mean-field potential, which is stronger than that from the SCGF approach. This effect is already present in SNM but becomes even stronger in PNM with a larger Fermi momentum, where the nucleon occupation probability is more affected by the momentum dependence of the nucleon mean-field potential. For the occupation probability from ImMDI-GF, it is similar to that from Sk$\chi$m$^*$  for SNM but closer to the SCGF results for PNM, since the correlation effects become less important in PNM. These similarities can be understood from the behaviors of the single-particle potentials in Figs.~\ref{U_SNM} and \ref{U_PNM} given in Sec.~\ref{MFsec}.

\subsection{Entropy and heat capacity}

The nucleon occupation probability $n(p)$ discussed in the previous subsection allows us to calculate the entropy per nucleon $S$ according to
\begin{equation}
S = -\frac{\nu}{\rho} \int \frac{d^3p}{(2\pi)^3} \left[ n\ln n + (1-n)\ln(1-n) \right],
\label{entropyqp}
\end{equation}
and the heat capacity at constant volume through the relation
\begin{equation}
c_v = T \left(\frac{\partial S}{\partial T}\right)_{\delta,\rho}.
\end{equation}
We must point out that within the SCGF method the entropy is obtained following the Luttinger-Ward formalism and it comprises both terms describing the quasi-particle behavior of the system, such as Eq.~\eqref{entropyqp}, and those related to fragmentation effects due to correlations in the many-body system~\cite{Rios2009,Carbone2018}.

\begin{figure}[htbp]
	\includegraphics[scale=0.35]{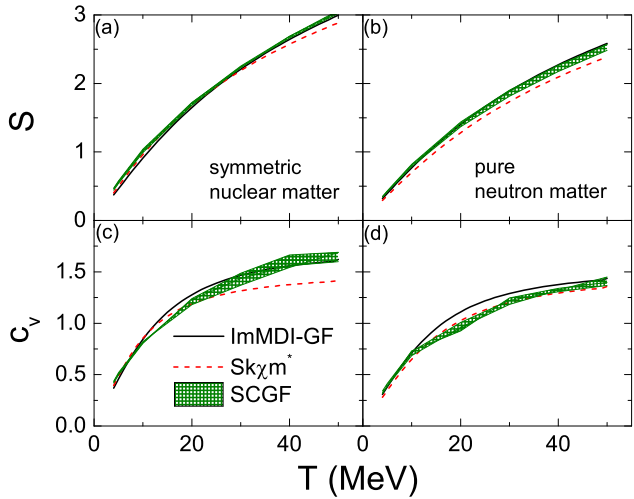}
	\caption{(Color	online) Entropy per nucleon (upper) and heat capacity at constant volume (lower) at $\rho_0=0.16$ fm$^{-3}$ as a function of the temperature in symmetric nuclear matter (left) and pure neutron matter (right) from the ImMDI-GF, Sk$\chi$m$^*$, and SCGF calculations.}\label{Scv}
\end{figure}

Figure~\ref{Scv} compares the entropy per nucleon as well as the heat capacity at constant volume at $\rho_0=0.16$ fm$^{-3}$ as a function of the temperature in SNM and PNM from the ImMDI-GF, Sk$\chi$m$^*$, and SCGF calculations. One can in principle relate the behavior of the entropy to that of the occupation probability in Fig.~\ref{f}. In the case of SNM, the more diffusive nucleon distribution in the SCGF approach due to many-body correlations leads to a slightly larger entropy per nucleon than those from ImMDI-GF and Sk$\chi$m$^*$. Since the correlation effect is less strong in PNM, the occupation probability and the entropy per nucleon from the SCGF approach are both similar to those from ImMDI-GF, while the entropy per nucleon from Sk$\chi$m$^*$ is lower as a result of the sharper nucleon momentum distribution. As a measurable quantity, the heat capacity at constant volume from the SCGF approach is mostly similar to that from ImMDI-GF for SNM, while it stands closer to the Sk$\chi$m$^*$ results for PNM, which gives a relatively smaller heat capacity especially at higher temperatures.

\subsection{Kinetic and potential energy contributions to the nuclear matter EOS}

\begin{figure}[htbp]
	\includegraphics[scale=0.35]{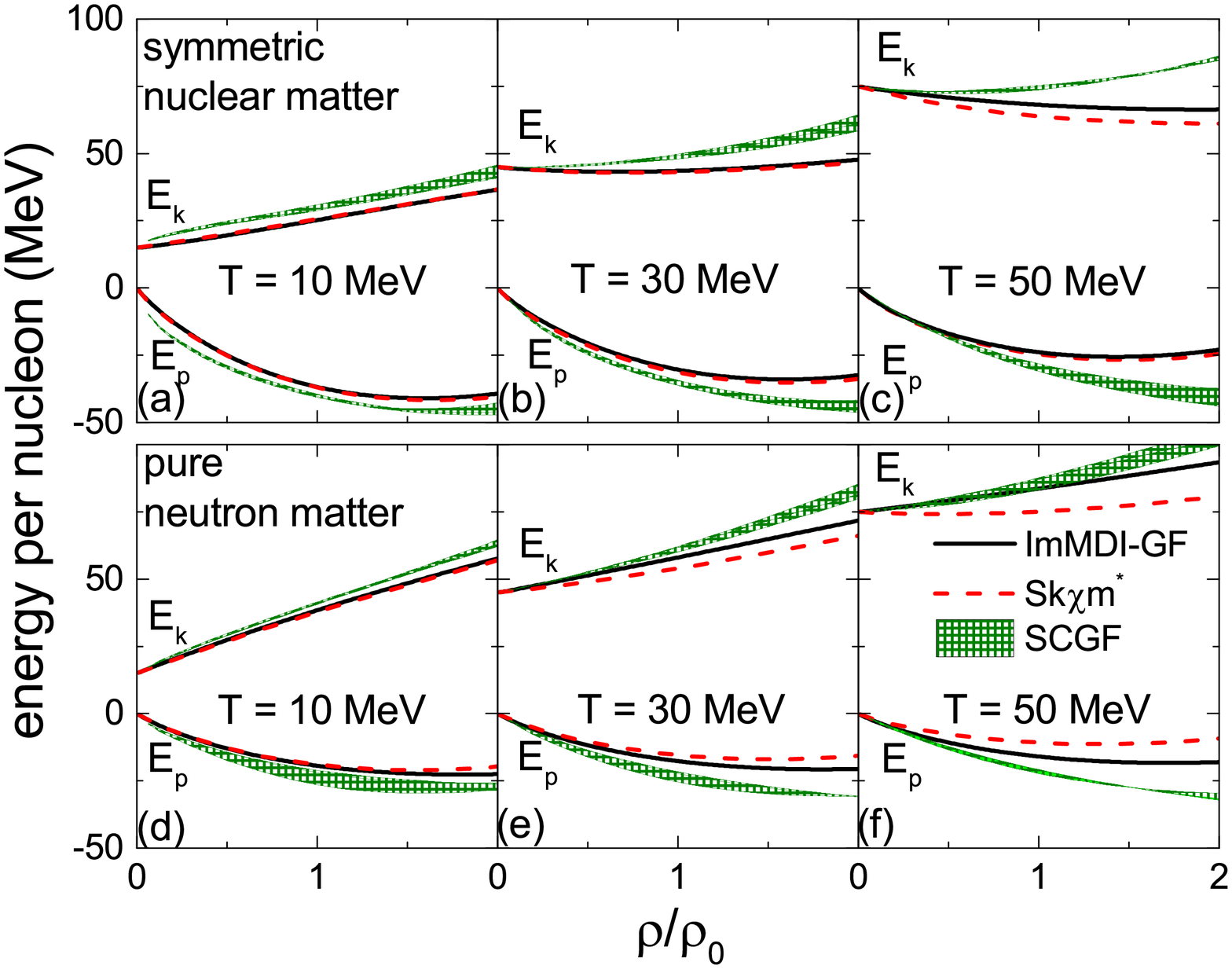}
	\caption{(Color	online) Kinetic energy $E_k$ and potential energy $E_p$ per nucleon as a function of reduced nucleon density for symmetric nuclear matter (upper) and pure neutron matter (lower) at various temperatures from the ImMDI-GF, Sk$\chi$m$^*$, and SCGF calculations.}\label{Ekp}
\end{figure}

The kinetic energy contribution, which is uniquely determined by the nucleon occupation probability as indicated in Eq.~(\ref{kinetic}), and the potential energy contribution to the EOS obtained from different approaches, are compared in Fig.~\ref{Ekp}. Since ImMDI-GF and Sk$\chi$m$^*$ are constructed from fitting similar nuclear EOSs at zero temperature, and they also have same nucleon occupation probabilities at zero temperature, the kinetic energy and the potential energy contribution to the EOS of cold nuclear matter from the two effective interactions are almost identical. The kinetic energy contributions to the EOS from ImMDI-GF and Sk$\chi$m$^*$ start to deviate as temperature increases, especially for PNM, with Sk$\chi$m$^*$ always giving smaller values. The fact that the kinetic energy contributions in SNM and PNM from effective interactions based on the Hartree-Fock calculations are always below those from SCGF is consistent with the behavior of nucleon occupation probabilities shown in Fig.~\ref{f}, where the SCGF always gives a larger population of high-momentum states and thus a larger kinetic energy, as a result of correlation effects. Deviations between the results on kinetic energy contributions in both SNM and PNM from ImMDI-GF and Sk$\chi$m$^*$ increase with both increasing density and temperature as a result of the different momentum dependence in their mean-field potentials.

For the potential energy contribution, which depends on both the density and the nucleon occupation probability, results from the ImMDI-GF and Sk$\chi$m$^*$ are in good agreement at all temperatures for SNM, while they start to deviate for PNM as temperature increases. This is again consistent with the results for the nucleon occupation probability shown in Fig.~\ref{f}. The potential energy contribution from the SCGF approach in both SNM and PNM is, however, always lower compared to that from ImMDI-GF and Sk$\chi$m$^*$ due to its larger nucleon occupation probability at high momenta as shown in Fig.~\ref{f}.

\subsection{EOSs of symmetric and pure neutron matter}

\begin{figure}[htbp]
	\includegraphics[scale=0.35]{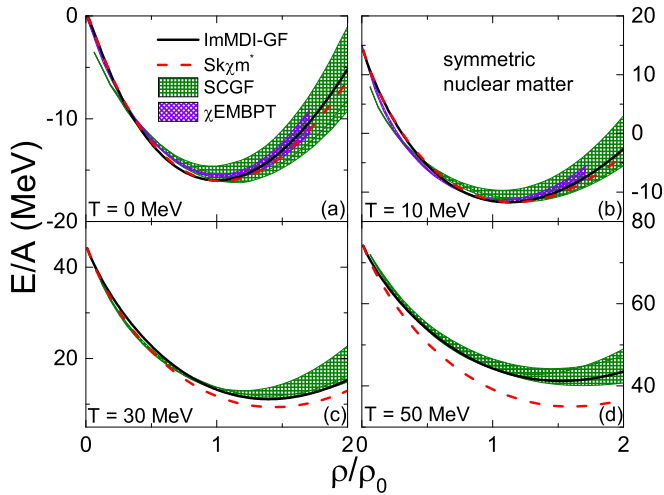}
	\caption{(Color	online) Total energy per nucleon as a function of reduced nucleon density for symmetric nuclear matter at various temperatures from ImMDI-GF and Sk$\chi$m$^*$ compared with results from the SCGF approach. The uncertainty bands for the $\chi$EMBPT in panels (a) and (b) are due to the use of two different n3lo414 and n3lo450 forces~\cite{Wel15}. }\label{EOS_SNM}
\end{figure}

\begin{figure}[htpb]
	\includegraphics[scale=0.35]{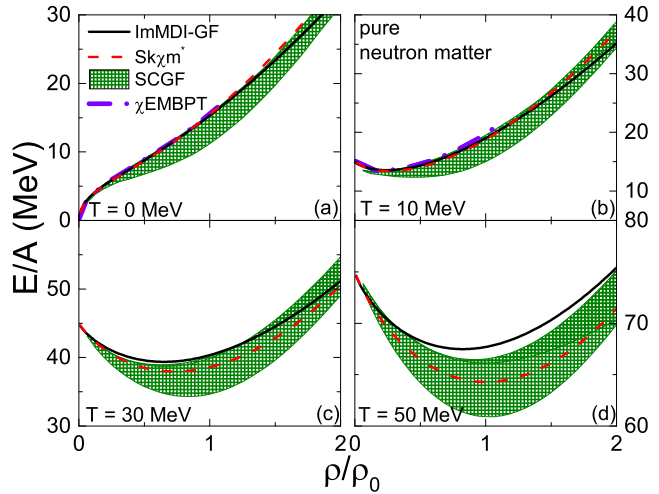}
	\caption{(Color	online) Same as Fig.~\ref{EOS_SNM} but for pure neutron matter. }\label{EOS_PNM}
\end{figure}

The density dependence of the total energy per nucleon for SNM and PNM from ImMDI-GF and Sk$\chi$m$^*$ are compared with results from the SCGF approach in Figs.~\ref{EOS_SNM} and \ref{EOS_PNM}, respectively. It is seen that the uncertainty in the SCGF results becomes larger at higher nucleon densities because of the different high-momentum cutoffs and three-body forces, particularly for the EOS of PNM. The EOSs from ImMDI-GF and Sk$\chi$m$^*$ are similar for cold and low-temperature SNM and PNM. Except for small deviations at very low densities, the EOSs from ImMDI-GF and Sk$\chi$m$^*$ are within the SCGF uncertainty band (see panels (a) and (b) in Figs.~\ref{EOS_SNM} and \ref{EOS_PNM}). However, results start to deviate at higher temperatures, with the EOS of SNM from ImMDI-GF remaining within the uncertainty band of SCGF but that from Sk$\chi$m$^*$ becoming slightly lower. For the EOS of PNM, ImMDI-GF gives slightly larger values at very high temperatures, while that from Sk$\chi$m$^*$ is within the uncertainty band of SCGF. These results can be partially understood from the relative contributions of the kinetic energy and the potential energy to the EOS, as shown in Fig.~\ref{Ekp}. Also shown in panels (a) and (b) for both SNM and PNM are results from $\chi$EMBPT calculations using n3lo414 and n3lo450 chiral forces, which are taken from Figs.~1 and 2 of Ref.~\cite{Wel15}. The uncertainty band in the latter approach for SNM is smaller than that given by SCGF due to a smaller range of variation in the high-momentum cutoff and similar three-body forces used in the two potentials. For PNM at low temperatures, results from the $\chi$EMBPT using n3lo414 and n3lo450 forces give almost identical EOS, due to reduced regulator dependence in the three-body forces~\cite{Coraggio2013}. The EOSs of both SNM and PNM from the $\chi$EMBPT are well reproduced by ImMDI-GF and Sk$\chi$m$^*$ at both $T=0$ and 10 MeV.

%\subsection{Comparison of the mean-field potential}
\subsection{Nucleon mean-field potentials in symmetric and pure neutron matter}
\label{MFsec}

\begin{figure}[t]
	\includegraphics[scale=0.35]{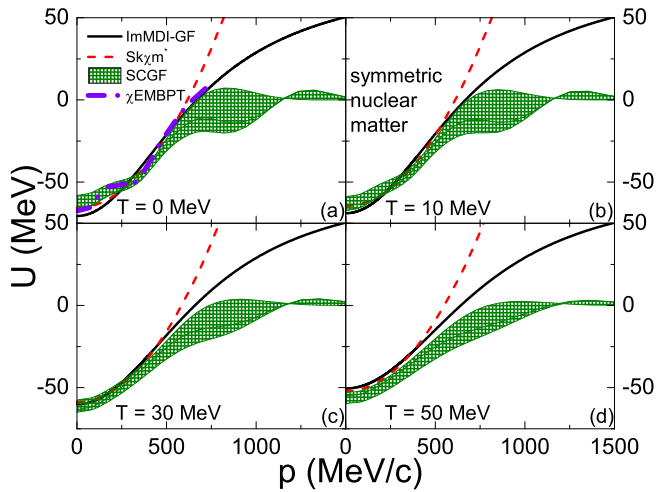}
	\caption{(Color	online) Mean-field potential at $\rho_0=0.16$ fm$^{-3}$ as a function of nucleon momentum in symmetric nuclear matter at various temperatures from ImMDI-GF and Sk$\chi$m$^*$ compared with results from the SCGF approach. Result at $T=0$ MeV from the $\chi$EMBPT using the n3lo450 force~\cite{Hol13} is also shown for comparison.}\label{U_SNM}
\end{figure}

\begin{figure}[t]
	\includegraphics[scale=0.35]{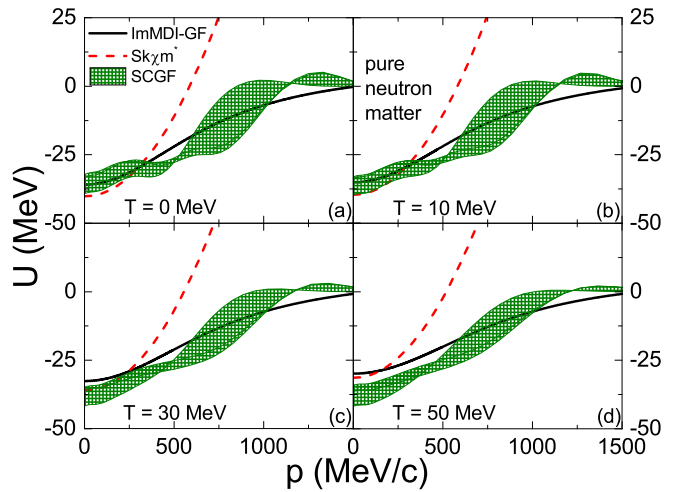}
	\caption{(Color	online) Same as Fig.~\ref{U_SNM} but for pure neutron matter.}\label{U_PNM}
\end{figure}

In transport simulations of intermediate-energy heavy-ion collisions, the direct input is the mean-field potential instead of the EOS. The temperature dependence of the mean-field potential is thus important in determining the evolution of the hot nuclear matter produced in these collisions. We compare in this subsection the momentum dependence of the mean-field potential at $\rho_0$ obtained from ImMDI-GF and Sk$\chi$m$^*$ with that from the SCGF approach in Figs.~\ref{U_SNM} and \ref{U_PNM} for SNM and PNM, respectively. For the SCGF approach, the mean-field potential given in Eq.~\eqref{epsilon} is obtained from the on-shell part of the real self-energy. The mean-field potentials from the SCGF approach in all these different cases are seen to always approach zero at nucleon momenta $\sim1000$ MeV/c. This is due to the high momentum cutoff in the regulator functions used in constructing these chiral forces~\cite{Ent03}. For the case of effective interactions, the mean-field potential at the saturation density is generally fitted from the energy dependence of nuclear optical potentials in elastic nucleon-nucleus scatterings~\cite{Jeu76,Joh87}. With its proper isoscalar effective masses as well as the mean-field potential $U_0^\infty$ at saturation density and infinite large nucleon momentum, the ImMDI model can well reproduce the momentum dependence of the optical potential extracted by Hama et al.~\cite{Ham90,Coo93} from the proton-nucleus scattering data as shown in Fig.~1 of Ref.~\cite{Xu15}.

For SNM at low temperatures, the mean-field potentials from both ImMDI-GF and Sk$\chi$m$^*$ are consistent with results from the SCGF approach up to $p=500$ MeV/c, while for PNM these results start to deviate already below $p=500$ MeV/c~\cite{footnote1}. With its isovector effective mass adjusted to be about $m_v^*=0.59m$, the ImMDI-GF interaction gives mean-field potentials in PNM that are consistent with those from the SCGF approach even at higher momenta. This is different for the mean-field potentials from Sk$\chi$m$^*$, which are seen to increase quadratically with nucleon momentum and become more repulsive around $p=500$ MeV/c in SNM. This is especially so for PNM where the deviations appear already at lower momenta. This is understandable, as discussed in Sec.~\ref{EIHF}, since the momentum dependence of the mean-field potential in the SHF model is only valid at lower nucleon momenta. Both ImMDI-GF and Sk$\chi$m$^*$ reproduce very well the mean-field potential at low momenta from the $\chi$EMBPT using the n3lo450 force at $T=0$ MeV for SNM~\cite{Hol13}. The temperature effect on the mean-field potentials from ImMDI-GF and Sk$\chi$m$^*$ is, on the other hand, stronger than that from the SCGF approach, especially at lower nucleon momenta. It is remarkable that the momentum dependence of the mean-field potentials from the SCGF approach for both SNM and PNM are reproduced reasonably well by ImMDI-GF at all considered temperatures.

%(mean-field potential at twice normal density?)

\section{Summary}
\label{summary}

To study if the commonly used nucleon-nucleon effective interactions, which are usually constructed from fitting the properties of cold nuclear matter and of finite nuclei, can properly describe nuclear matter at finite temperatures, we have used an improved isospin- and momentum-dependent interaction ImMDI-GF and the recently constructed Skyrme interaction Sk$\chi$m* to evaluate the nucleon occupation probabilities, the equations of state, and the mean-field potentials in symmetric nuclear matter and pure neutron matter at finite temperatures using the Hartree-Fock approach. These results have been compared with those from the microscopic self-consistent Green's function method and the chiral effective many-body perturbation theory using chiral nuclear forces. We have found significant differences between results from ImMDI-GF and Sk$\chi$m$^*$ for nuclear matter properties at high temperatures and also between results from these two models and those from the microscopic theories. The deviations seen in the nucleon occupation probabilities in these approaches have been understood from their different momentum dependence in the single-nucleon potential, which is strongly suppressed at high momenta in the microscopic calculations based on chiral forces compared to those from the effective interactions, especially for Sk$\chi$m$^*$ that has a quadratic momentum dependence. These differences in the nucleon momentum distributions have led to deviations in the kinetic energy contribution and also partially in the potential energy contribution to the nuclear equation of state. The energies per nucleon for symmetric nuclear matter and pure neutron matter from ImMDI-GF and Sk$\chi$m$^*$ are roughly consistent with those from the self-consistent Green's function approach, although the equation of state for symmetric nuclear matter from Sk$\chi$m* remains softer at higher temperatures compared to the other two approaches. Using ImMDI-GF in the Hartree-Fock calculation reproduces remarkably well the mean-field potential from the microscopic approaches at various temperatures for both symmetric nuclear matter and pure neutron matter. Our study thus shows that effective interactions with the correct momentum dependence in the mean-field potential, such as the one from ImMDI-GF, can properly describe the properties of hot dense nuclear matter and is thus suitable for use in transport models to extract the equation of state of cold nuclear matter, which is needed for describing the properties of neutron stars, from intermediate-energy heavy-ion collisions.

The SCGF approach using chiral forces with high-momentum cutoffs has, however, larger theoretical uncertainties at high densities.  Also, the mean-field potential at suprasaturation densities from effective interactions using the Hartree-Fock approach is based on the extrapolation from normal density and thus depends on the functional form used in the parametrization of its value at this density.  Efforts have been made in the past to constrain the momentum dependence of the isoscalar mean-field potential by comparing transport model results of collective flows in heavy-ion collisions at several hundred AMeV with the experimental data~\cite{Pan93,Zha94,Dan00}.
To constrain not only the temperature dependence but also the density dependence of the nuclear mean-field potential, both well-defined effective interaction functionals and reliable transport models are needed to study experimental observables that are sensitive to the momentum dependence of nucleon mean-field potential~\cite{Xu16,Xu19}, which is, however, beyond the scope of present study.

\begin{acknowledgments}
We thank Jeremy Holt for providing results from the many-body perturbation theory using nuclear chiral forces. J.X. acknowledges support from the Major State Basic Research Development Program (973 Program) of China under Contract No. 2015CB856904 and the National Natural Science Foundation of China under Grant No. 11421505. C.M.K. acknowledges support from the US Department of Energy under Contract No. DE-SC0015266 and the Welch Foundation under Grant No. A-1358.
\end{acknowledgments}

\end{document}